\documentclass{PoS}

\usepackage{amsmath}
\usepackage{epsfig}
\usepackage{amssymb}
\usepackage{bbold}
\usepackage{braket}
\input{epsf}

\def\bea#1\eea{\begin{align}#1\end{align}}
\newcommand{\nnu}{\nonumber\\}
\newcommand{\bef}{\begin{figure}[htb]\centering}
\newcommand{\eef}{\end{figure}}

\def\be{\begin{equation}}
\def\ee{\end{equation}}
\def\beq{\begin{equation}}
\def\eeq{\end{equation}}
\def\bea{\begin{eqnarray}}
\def\eea{\end{eqnarray}}
\def\ba{\begin{eqnarray}}
\def\ea{\end{eqnarray}}
\def\eeq{\end{equation}}
\def\beeq{\begin{eqnarray}}
\def\eeeq{\end{eqnarray}}
\def\beqa{\begin{eqnarray}}
\def\eeqa{\end{eqnarray}}

\def\to{\rightarrow}

\title{A Study of Quasi-parton Distribution Functions in the Diquark Spectator Model}

\ShortTitle{Quasi-PDFs form the Diquark Spectator Model}

\author{Leonard Gamberg\\
          Division of Science,  Penn State Berks, Reading, PA 19610, USA\\
          E-mail: \email{lpg10@psu.edu}}

\author{Zhong-Bo Kang\\
        Theoretical Division, MS B283, Los Alamos National Laboratory, Los Alamos, NM 87545, USA\\
        E-mail: \email{zkang@lanl.gov}}

\author{\speaker{Ivan Vitev}\\
        Theoretical Division, MS B283, Los Alamos National Laboratory, Los Alamos, NM 87545, USA\\
        E-mail: \email{ivitev@lanl.gov}}

\author{Hongxi Xing\\
        Theoretical Division, MS B283, Los Alamos National Laboratory, Los Alamos, NM 87545, USA\\
        E-mail: \email{hxing@lanl.gov}}

\abstract{To facilitate lattice QCD calculations of nucleon structute, a set of quasi-parton distributions were
recently introduced. These quasi-PDFs were shown to reduce to standard PDFs when the nucleon is boosted
to high energies, $P_z\rightarrow \infty$. Since taking such limit is not feasible in lattice simulations, it is essential to provide guidance for what values of $P_z$ the quasi-PDFs are good approximations of standard PDFs. Within the framework of the spectator diquark model, we  evaluate both the up and down quarks' quasi-PDFs and standard PDFs for all leading-twist distributions (unpolarized distribution $f_1$, helicity distribution $g_1$, and transversity distribution $h_1$). We find that, for intermediate parton momentum fractions $x$, quasi-PDFs are good approximations to standard PDFs (within $20-30\%$) when $P_z\gtrsim 1.5-2$ GeV. On the other hand, for large $x\sim 1$  much larger $P_z > 4$~GeV is necessary to obtain a satisfactory agreement between the two sets. We further find that the Soffer positivity bound does not hold in general for quasi-PDFs.   }
          
\FullConference{QCD Evolution 2015\\
		          May 26-30, 2015\\
		          Jefferson Lab (JLAB), Newport News Virginia, USA}

\begin{document}

\section{Introduction}

In recent years, evaluation of parton distribution functions (PDFs), fundamental non-perturbative ingredients of
the QCD factorizaton approach, has been attempted in lattice QCD~\cite{Deka:2008xr,Alexandrou:2014yha,Hagler:2009ni,Musch:2011er}.
Since PDFs are defined as the non-local light-cone correlations which involve the real Minkowski time, the traditional lattice QCD approach does not allow one to compute the PDFs {\em directly}~\cite{Ji:2013dva}; one can only calculate the lower moments of the PDFs, which are matrix elements of local operators~\cite{Deka:2008xr,Alexandrou:2014yha}. Recently, new methods have been proposed \cite{Ji:2013dva,Ma:2014jla} to evaluate  PDFs on the lattice in terms of so-called  quasi-PDFs,  which are
 defined as matrix elements of equal-time spatial correlators. 
These quasi-PDFs can be computed directly on the lattice~\cite{Lin:2014zya,Alexandrou:2014pna,Liu:2015nva} and should reduce to the standard PDFs when the proton's momentum $P_z\to \infty$. While in practice the proton momentum on the lattice can never become infinite, one can only  hopefully  access finite but large enough momenta on the lattice to carry out relevant QCD simulations. Here, we present guidance regarding the magnitude of the proton momentum based upon the comparision of PDFs and quasi-PDFs in the spectator diquark model \cite{Gamberg:2014zwa}.

\section{Overview and definitions of standard PDFs and quasi-PDFs: }

We consider a nucleon of mass $M$ moving in the $z$-direction, with the momentum $P^\mu$ given by
\be
P^\mu=(P_0, 0_\perp, P_z) \equiv [P^+, P^-, 0_\perp].
\ee
Here and throughout the paper we use $(v_0, v_\perp, v_z)$ and $[v^+, v^-, v_\perp]$ to represent Minkowski and light-cone components for any four-vector $v^\mu$ respectively, with light-cone variables $v^{\pm} = (v_0\pm v_z)/\sqrt{2}$. We thus have
\be
P^- = \frac{M^2}{2P^+}, \qquad
P_0 = \sqrt{P_z^2 + M^2} \equiv P_z \delta, \quad {\rm with }  \quad \delta=\sqrt{1+\frac{M^2}{P_z^2}}.
\ee

For the helicity distribution $g_1$  and the transversity distribution $h_1$ we also have to consider the nucleon with either longitudinal or transverse polarization. For  pure longitudinal polarization, the polarization vector $S_L^\mu$ is given by
\be
S_L^\mu =\frac{1}{M}\left(P_z, 0_\perp, P_0\right) \equiv \frac{1}{M}\left[P^+, -P^-, 0_\perp\right].
\label{eq:sl}
\ee
On the other hand, for  pure transverse polarization, we have the polarization vector $S_T^\mu$
\be
S_T^\mu = (0, \vec{S}_\perp, 0) \equiv [0^+, 0^-, \vec{S}_\perp].
\label{eq:st}
\ee
The polarization vectors satisfy the conditions  $P\cdot S_L = P\cdot S_T = 0$,  and $S_L^2 = -1$ and $S_T^2 = -\vec{S}_\perp^2 = -1$.

The three leading-twist standard collinear PDFs are defined on the light-cone with the following operator expressions~\cite{Brock:1993sz}
\begin{eqnarray}
f_1(x) &= &\int \frac{d\xi^-}{4\pi} e^{-i \xi^-k^+}  \langle P|\overline{\psi}(\xi^-) \gamma^+ U_n[\xi^-,0] \psi(0) |P\rangle,
\\
g_1(x) &=& \int \frac{d\xi^-}{4\pi} e^{-i \xi^-k^+}  \langle PS|\overline{\psi}(\xi^-) \gamma^+\gamma_5 U_n[\xi^-,0] \psi(0) |PS\rangle,
\\
h_1(x) &=& \int \frac{d\xi^-}{4\pi} e^{-i \xi^-k^+}  \langle PS|\overline{\psi}(\xi^-) \gamma^+\gamma_5 \gamma\cdot S_T U_n[\xi^-,0] \psi(0) |PS\rangle,
\end{eqnarray} 
with $x=k^+/P^+$. We define the light-cone vector $n^\mu = [0^+, 1^-, 0_\perp]$ with $n^2 = 0$ and $n\cdot v=v^+$ for any four-vector $v^\mu$, and  
the gauge link $U_n[\xi^-,0]$ along the light-cone direction specified by $n$ is given by
\bea
U_n[\xi^-,0] = \exp\left(-ig\int^{\xi^-}_0 d\eta^- A^+(\eta^-) \right).
\eea
On the other hand, the quasi-PDFs introduced by Ji \cite{Ji:2013dva} are equal-time spatial correlations along the $z$-direction, and have the following operator definitions
\begin{eqnarray}
\tilde f_1(x, P_z) &=& \int \frac{d\xi_z}{4\pi} e^{ -i \xi_z k_z}  \langle P|\overline{\psi}(\xi_z) \gamma_z U_{n_z}[\xi_z,0] \psi(0) |P\rangle,
\\
\tilde g_1(x, P_z) &=& \int \frac{d\xi_z}{4\pi} e^{-i \xi_z k_z}  \langle PS|\overline{\psi}(\xi_z) \gamma_z\gamma_5 U_{n_z}[\xi_z,0] \psi(0) |PS\rangle,
\\
\tilde h_1(x, P_z) &=& \int \frac{d\xi_z}{4\pi} e^{-i \xi_z k_z}  \langle PS|\overline{\psi}(\xi_z) \gamma_z \gamma_5 \gamma\cdot S_T U_{n_z}[\xi_z,0] \psi(0) |PS\rangle,
\end{eqnarray}
where $n_z^\mu = (0, 0_\perp, 1)$ with $n_z^2 = -1$ and $n_z \cdot v = -v_z$ for any four-vector $v^\mu$, where now the gauge link $U_{n_z}[\xi_z,0]$ is 
along the direction of $n_z$ and is  given by
\be
U_{n_z}[\xi_z,0] = \exp\left(-ig\int^{\xi_z}_0 d\eta_z A_z (\eta_z)   \right).
\ee

\section{The spectator diquark model}
The spectator diquark model of the nucleon has been described in great detail~\cite{Jakob:1997wg, Gamberg:2007wm, Bacchetta:2008af, Kang:2010hg}. Here, we present a brief overview. In the spectator diquark model, the PDFs which are traces of the quark-quark correlation functions as defined in the last section are evaluated in the spectator approximation. In this framework a sum over a complete set of intermediate on-shell states, 
 $I=\sum_{X}|X\rangle\langle X|$, is  inserted into the operator definition of PDFs, and truncated to single on-shell diquark spectator states with $X$ being 
either spin 0 (scalar diquark) or spin 1 (axial-vector diquark). The quark-quark correlation function is then obtained as the cut tree level amplitude for nucleon $N\rightarrow q\, + \, X$ where $X={\{s,a\}}$. With such an approximation, the nucleon is composed of a constituent quark of mass $m$ and a spectator scalar (axial-vector) diquark with mass $M_s$ ($M_a$). 
\bef
\psfig{file=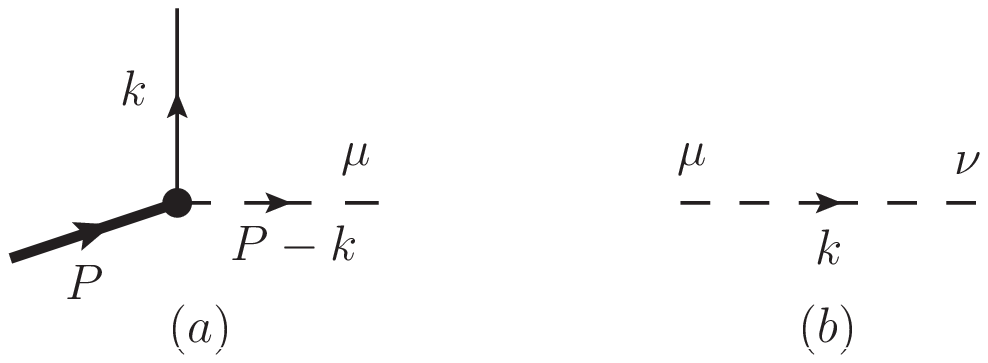, width=3in}
\caption{Feynman rules in the spectator diquark model: (a) vertex representing the interaction between the quark, the nucleon, and the diquark, (b) the diquark propagator.}
\label{fig:feynmanrule}
\eef
The interaction between the nucleon, the quark, and the diquark is given by the following Feynman rules for the vertex in Fig.~\ref{fig:feynmanrule}(a),
\be
\mbox{scalar diquark:}  \quad  ig_s \mathcal{I}_s(k^2), \qquad
\mbox{axial-vector diquark:}  \quad i\frac{g_a}{\sqrt{2}} \gamma^\mu \gamma_5 \mathcal{I}_a(k^2),
\label{eq:axial-vertex}
\ee
where following~\cite{Gamberg:2007wm,Bacchetta:2008af, Kang:2010hg}, we have introduced suitable form factors $\mathcal{I}_{s,a}(k^2)$ as a function of $k^2$ - the invariant mass of the constituent quark. For our numerical calculations below, 
we adopt the fitted parameters in~\cite{Bacchetta:2008af} and use
 the  dipolar form factors,
\begin{equation}
\mathcal{I}_s(k^2) = \frac{k^2-m^2}{\left(k^2 - \Lambda_s^2\right)^2},
\qquad
\mathcal{I}_a(k^2) = \frac{k^2-m^2}{\left(k^2 - \Lambda_a^2\right)^2},
\label{eq:form}
\end{equation}
where $\Lambda_{s,a}$ are the appropriate cutoffs, to be considered as free parameters of the model together with the diquark masses $M_{s,a}$, and the couplings $g_{s,a}$. Further, 
 the propagators of the scalar diquark and the axial-vector diquark as shown in Fig.~\ref{fig:feynmanrule}(b) are given by the expressions, 
\begin{equation}
\mbox{scalar diquark:}  \quad  \frac{i}{k^2-M_s^2}, \qquad
\mbox{axial-vector diquark:}  \quad  \frac{i}{k^2-M_a^2}d^{\mu\nu}(k, n),
\label{eq:axial-propagator}
\end{equation}
where for the standard light-cone PDFs with $n^2=0$, we have~\cite{Bacchetta:2008af, Kang:2010hg}
\be
d^{\mu\nu}(k, n)  = -g^{\mu\nu} + \frac{n^\mu k^\nu + n^\nu k^\mu}{n\cdot k} - \frac{k^2 n^\mu n^\nu}{\left(n\cdot k\right)^2},
\label{eq:axial-pol}
\ee
which satisfies $n_\mu d^{\mu\nu}(k, n) = k_\mu d^{\mu\nu}(k, n) = 0$. On the other hand, for the quasi-PDFs, since $n_z^2 =-1\neq 0$, we have a slightly different form for the polarization tensor $d^{\mu\nu}$ as
\be
d^{\mu\nu}(k, n_z) =  -g^{\mu\nu} + \frac{n_z\cdot k}{\left(n_z\cdot k\right)^2 - n_z^2 k^2} 
\left(n_z^\mu k^\nu + n_z^\nu k^\mu\right)
- \frac{1}{\left(n_z\cdot k\right)^2 - n_z^2 k^2} \left(k^2 n_z^\mu n_z^\nu + n_z^2 k^\mu k^\nu
\right),
\label{eq:axial-quasi-pol}
\ee
which also satisfies $n_{z\mu} d^{\mu\nu}(k, n_z) = k_{\mu} d^{\mu\nu}(k, n_z) = 0$.

\section{Standard PDFs and Quasi-PDFs in the spectator diquark model}

We give one detailed example of the calculation of standard PDFs and quasi-PDFs. 
In the scalar diquark model~\cite{Jakob:1997wg}, $f_1^s(x, k_\perp^2)$  is given by
\bea
f_1^s(x, k_\perp^2) &=& g_s^2 \int \frac{dk^+ dk^-}{(2\pi)^4} \frac{1}{2P^+} \delta\left(x - \frac{k^+}{P^+}\right){\rm Tr}\left[\gamma\cdot n \left(\gamma\cdot k+m \right) \frac{1}{2}\left(\gamma\cdot P+M\right)\left(\gamma\cdot k+m\right) \right]
\nnu
&&\times
\frac{1}{(k^2-m^2)^2} 2\pi \delta\left(\left(P-k\right)^2 - M_s^2\right) \left[\mathcal{I}_s(k^2)\right]^2,
\eea
where the superscript ``$s$'' in $f_1^s$ indicates that the diquark is a scalar, and $k_\perp$ is the quark transverse momentum~\cite{Bacchetta:2004jz,Bacchetta:2006tn}.  Eventually we obtain
\be
f_1^s(x,k_{\perp}^2=\frac{g_s^2}{(2\pi)^3}\frac{(1-x)[k_{\perp}^2+(m+xM)^2]}
{2\left[k_{\perp}^2+xM_s^2-x(1-x)M^2+(1-x)m^2\right]^2} \left[\mathcal{I}_s(k^2)\right]^2,
\label{eq:truef1}
\ee
where the invariant mass 
$k^2= - \frac{1}{1-x} \left[k_\perp^2+x M_s^2 - x(1-x)M^2 \right].$

Using the definition of the cut vertices for the quasi-PDFs, 
we  write the quasi-PDF $\tilde f_1^s(x, k_\perp^2, P_z)$ for the scalar diquark case as
\bea
\tilde f_1^s(x, k_\perp^2, P_z) &=& - g_s^2 \int \frac{dk_0 dk_z}{(2\pi)^4} \frac{1}{2P_z} \delta\left(x - \frac{k_z}{P_z}\right){\rm Tr}\left[\gamma\cdot n_z \left(\gamma\cdot k+m \right) \frac{1}{2}\left(\gamma\cdot P+M\right)\left(\gamma\cdot k+m\right) \right]
\nnu
&&\times
\frac{1}{(k^2-m^2)^2} 2\pi \delta\left(\left(P-k\right)^2 - M_s^2\right) \left[\mathcal{I}_s(k^2)\right]^2,
\eea
where we have used $\gamma_z = -\gamma\cdot n_z$. After some algebraic manipulation,
 the corresponding quasi-PDF $\tilde f_1^s(x, k_\perp^2, P_z)$ is given by
\be
{\tilde f}_1^s(x,k_{\perp}^2,P_z)=\frac{g_s^2}{(2\pi)^3} \frac{(2x-1)M^2+2x M m-M_s^2+m^2-2(1-x)^2(1-\rho_s\delta)P_z^2}
{2\rho_s(1-x)\left[2(1-x)(1-\rho_s\delta)P_z^2+M^2+M_s^2-m^2\right]^2} \left[\mathcal{I}_s(k^2)\right]^2,
\label{eq:tilde-f1}
\ee
where  for the quasi-PDFs 
$k^2 = 2(1-x)(1-\rho_s\delta)P_z^2+ M^2+M_s^2$.

We now study what happens to the quasi-PDF $\tilde f_1^s(x, k_\perp^2, P_z)$ 
in the limit  of $P_z\to \infty$. 
Approximating  $\rho_s$ and $\delta$ to $\mathcal{O}(M^2/P_z^2)$
\be
\label{eq:rho_s}
\rho_s  \approx 1+ \frac{k_{\perp}^2+M_s^2}{2(1-x)^2P_z^2},
\qquad
\delta   \approx 1+\frac{M^2}{2P_z^2},
\qquad {\rm and} \quad
(1-\rho_s\delta) P_z^2 \approx - \frac{k_{\perp}^2+M_s^2}{2(1-x)^2} - \frac{M^2}{2}.
\ee
Substituting this expression into the equations above we  find that
\be
\tilde f_1^s(x, k_\perp^2, P_z\to \infty) = f_1^s(x, k_\perp^2)\, .
\ee
Thus, the quasi-PDF reduces to the standard PDF $f_1^s(x, k_\perp^2)$ as  $P_z\to\infty$ limit~\footnote{Though obvious, it is worthwhile emphasizing that this conclusion is independent of the fact whether one has the form factor $\mathcal{I}_s(k^2)$ in the spectator diquark model.}. This simply verifies the leading order matching calculations carried out in \cite{Ji:2013dva,Ma:2014jla}. 

The approximation in Eq.~\eqref{eq:rho_s} we are using above seems quite reasonable. However, it is important to emphasize that such an approximation only holds when $(1-x)^2 \sim \mathcal O(1)$. When we are studying the quasi-PDFs in the very large $x\sim 1$ region, the large $P_z$ expansion used in Eq.~\eqref{eq:rho_s} breaks down, in which case the quasi-PDFs can deviate substantially 
from the standard PDFs. Such a breakdown is directly related to the existence of the factor $(1-x)^2P_z^2$ in our calculation, which is traced back to the on-shell condition of the diquark. Since such an on-shell condition is fairly generic~\cite{Xiong:2013bka}, 
we expect that it will be quite difficult for the quasi-PDFs to approach the standard PDFs in the large $x\sim 1$ region. In this case, one has to boost the proton to much larger $P_z$.  We will further illustrate this point in our numerical studies in the next section.

With the dipolar form factor $\mathcal{I}_s(k^2)$ given in Eq.~\eqref{eq:form}, one can further integrate $f_1^s(x, k_\perp^2)$ over $k_\perp^2$ to obtain the collinear distribution $f_1^s(x)$ as 
\be
f_1^s(x) = \int d^2k_\perp f_1^s(x,k_{\perp}^2) =  2\pi\int_0^{\infty} dk_\perp k_\perp f_1^s(x,k_{\perp}^2),
\ee
from which we obtain
\be
f_1^s(x) = \frac{g_s^2}{(2\pi)^2}\frac{\left[2(m+x M)^2 + L_s^2(\Lambda_s^2)\right] (1-x)^3}{24L_s^6(\Lambda_s^2)},
\ee
with $L_s^2(\Lambda_s^2)$ defined as
\be
L_s^2(\Lambda_s^2) \equiv xM_s^2 + (1-x) \Lambda_s^2 - x(1-x) M^2.
\ee
Now  let us consider  the quasi-PDF $\tilde f_1^s(x, P_z)$. We have
\be
\tilde f_1^s(x,P_z) = \int d^2k_\perp \tilde f_1^s(x,k_{\perp}^2, P_z) =  2\pi\int_0^{\infty} dk_\perp k_\perp \tilde f_1^s(x,k_{\perp}^2, P_z),
\ee
with $\tilde f_1^s(x,k_{\perp}^2, P_z)$ given by Eq.~\eqref{eq:tilde-f1}. Because of the complicated functional form for $\tilde f_1^s(x,k_{\perp}^2, P_z)$, we are not able to obtain a simple analytical expression for the collinear quasi-PDF $\tilde f_1^s(x, P_z)$ and will only present the numerical studies for the collinear quasi-PDFs in the next section. Here, it is important to 
emphasize that, since in the limit of $P_z\to \infty$, $\tilde f_1^s(x, k_\perp^2, P_z)$ reduces to $f_1^s(x, k_\perp^2)$ as we have shown above, the collinear counter-part $\tilde f_1^s(x, P_z)$ also reduces to the standard collinear PDF $f_1^s(x)$. 

We proceed to calculate the unintegrated  helicity and transversity distributions  $g_1^s(x,k_{\perp})$,   $g_1^a(x,k_{\perp})$, $h_1^s(x,k_{\perp})$,   $h_1^a(x,k_{\perp})$ for scalar
and axial diquarks. 
We also obtain the  quasi-helicity and quasi-transversity distributions  $\tilde g_1^s(x,k_{\perp},P_z)$,   $\tilde g_1^a(x,k_{\perp},P_z)$, $\tilde h_1^s(x,k_{\perp},P_z)$,   $\tilde h_1^a(x,k_{\perp},P_z)$. From the unintegrated
distributions one can proceed to get the integrated ones and all details are given in Ref.~\cite{Gamberg:2014zwa}.

\section{Phenomenological results \label{sec4}}

Following Ref.~\cite{Bacchetta:2008af}, the $u$-quark and $d$-quark unpolarized PDFs $f_1^{u,d}$ can be written as
\be
f_1^u = c_s^2 f_1^{u(s)} + c_a^2 f_1^{u(a)},
\qquad
f_1^d  = c_a'^{2} f_1^{d(a')},
\ee
that is, the $u$-quark receives contributions from both scalar and axial-vector diquark, while the $d$-quark only has the axial-vector diquark contribution. Here the superscript ``$s$'' represents the scalar diquark contribution, ``$a$'' corresponds to the axial-vector diquark which has isospin 0 (isoscalar $ud$-like system), and ``$a'$'' denotes the axial-vector diquark contribution which has isospin 1 (isovector $uu$-like system). Thus, we have the following 9 model parameters: $c_{s,a}$, $c_a'$, $M_{s,a}$, $M_a'$, $\Lambda_{s,a}$, and $\Lambda_a'$, as well as three couplings $g_s$, $g_a$, and $g_a'$. We use the same method specified in \cite{Bacchetta:2008af} to fix these three couplings: 
\be
\pi \int_0^1 dx \int_0^{\infty} dk_\perp^2 f_1^{q(X)}(x, k_\perp^2) = 1,
\ee
with $X=s, a, a'$. On the other hand, the other 9 model parameters are fixed through a global fitting of both $f_1^u(x)$, $f_1^d(x)$ at factorization scale $\mu^2=0.30 {\rm ~GeV}^2$ with ZEUS2002 PDFs~\cite{Chekanov:2002pv} and $g_1^u(x)$, $g_1^d(x)$ at $\mu^2=0.26 {\rm ~GeV}^2$  with GRSV2000~\cite{Gluck:2000dy} at leading order in~\cite{Bacchetta:2008af};  the fit is satisfactory and gives consistent shape and size of the standard PDFs. In the following, we simply use these fitted parameters in our numerical study.

\bef
\psfig{file=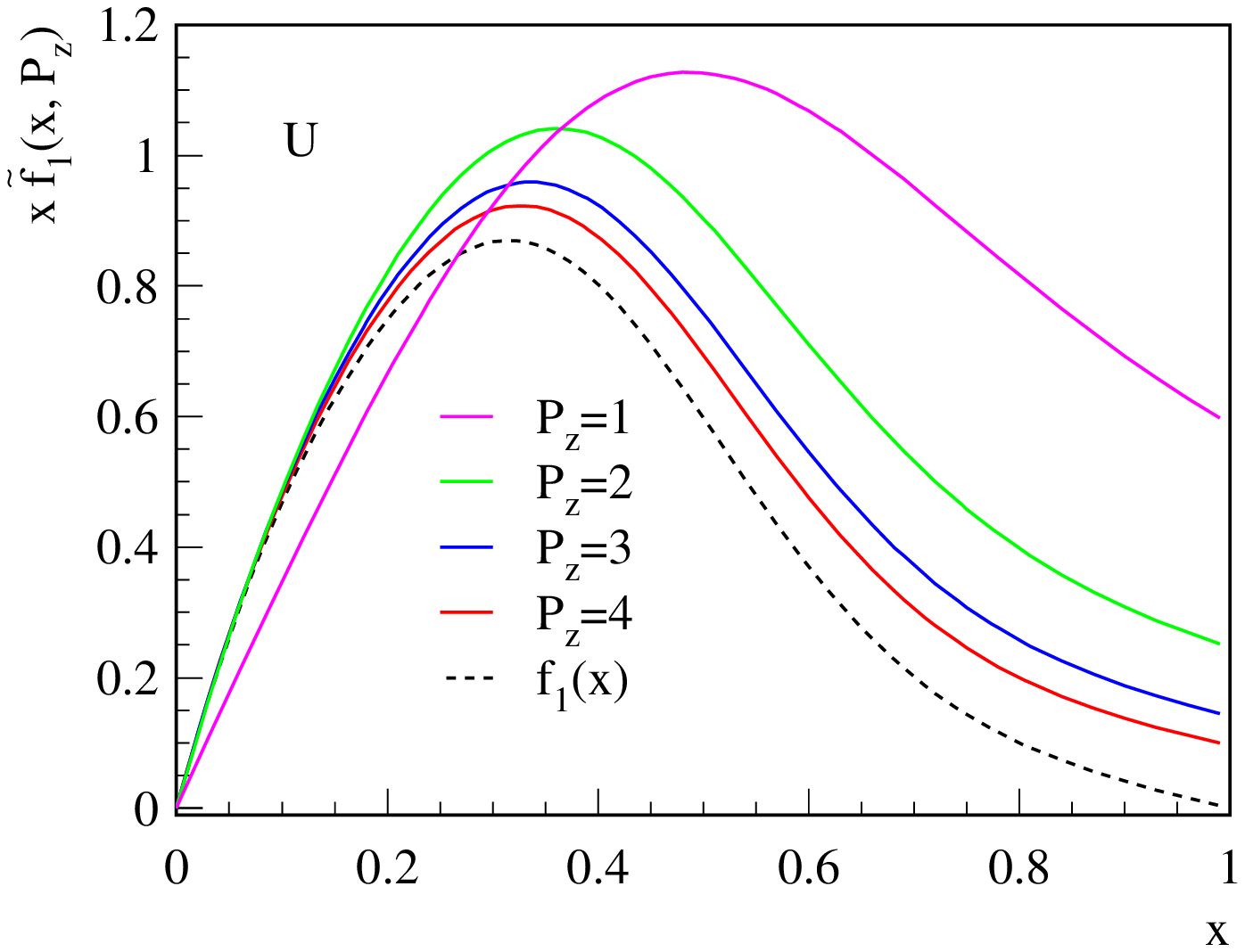, width=2.6in}
\hskip 0.2in
\psfig{file=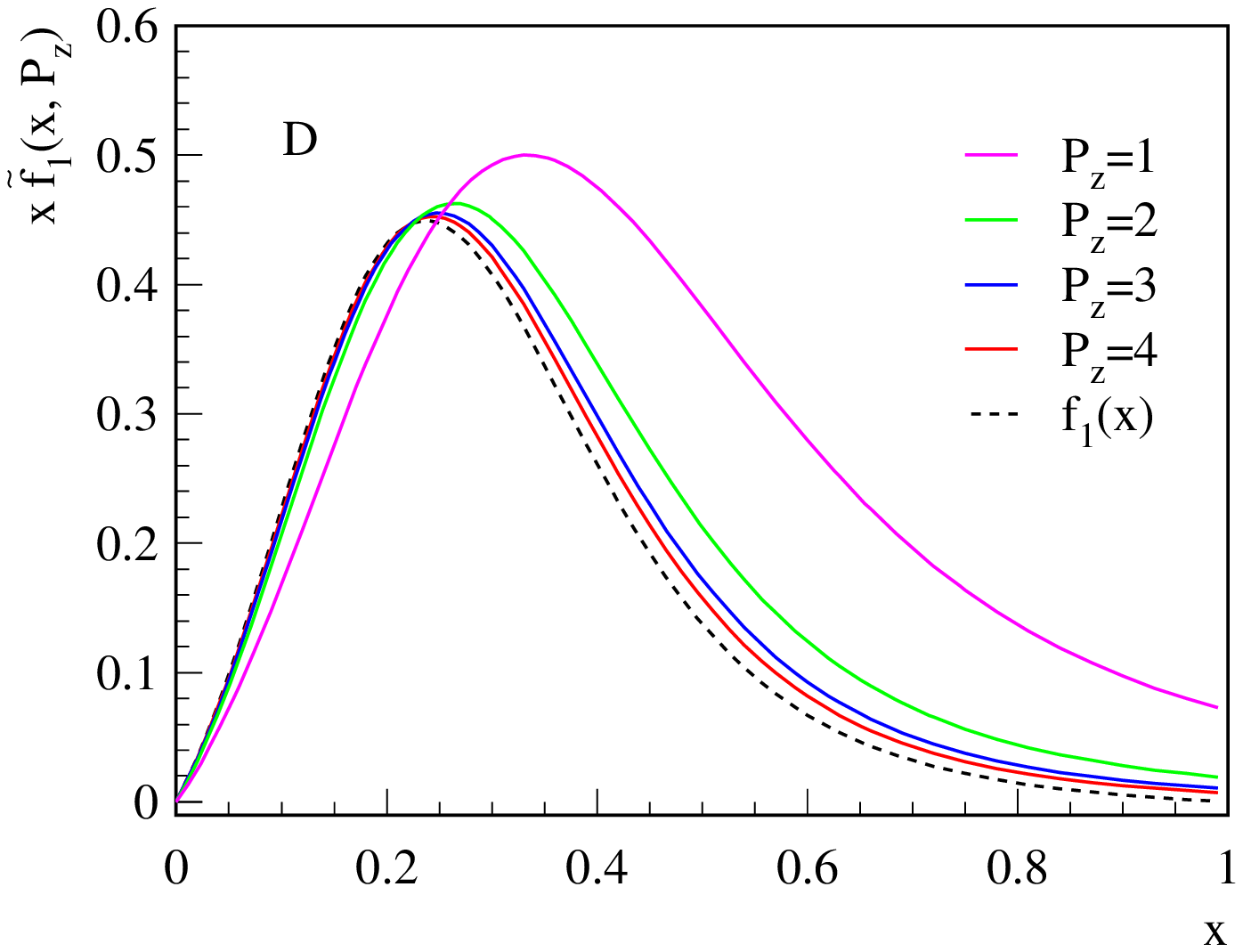, width=2.6in}
\caption{The unpolarized quasi-PDFs $x \tilde f_1(x, P_z)$ are plotted as a function of $x$ for $u$ (left) and $d$ (right) quark, respectively. Different lines are shown for $P_z=1$ GeV (purple), 2 GeV (green), 3 GeV (blue), and 4 GeV (red), respectively. The standard PDF $f_1(x)$ (black dashed) is also shown for comparison.}
\label{fig:f1-x}
\eef
\bef
\psfig{file=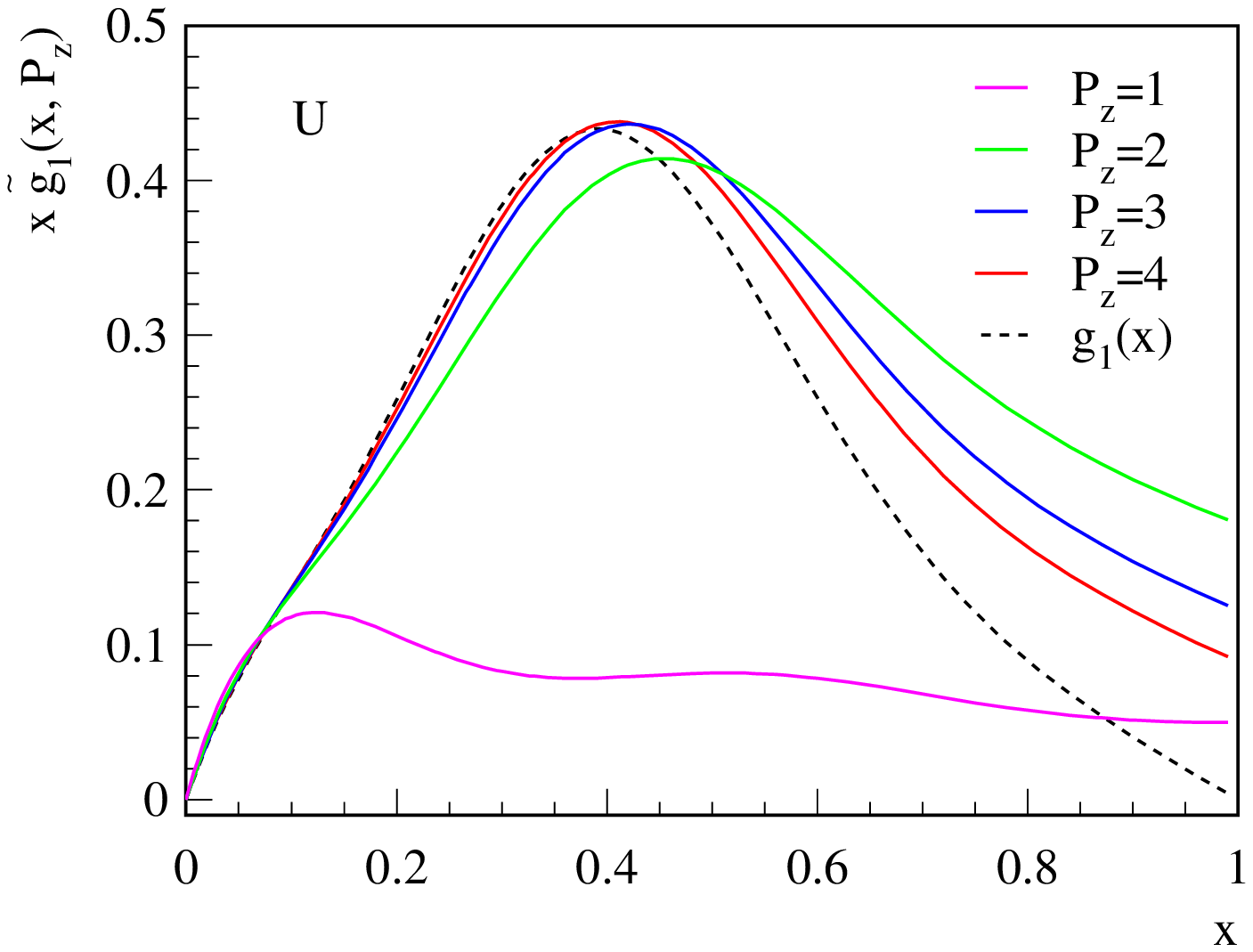, width=2.6in}
\hskip 0.2in
\psfig{file=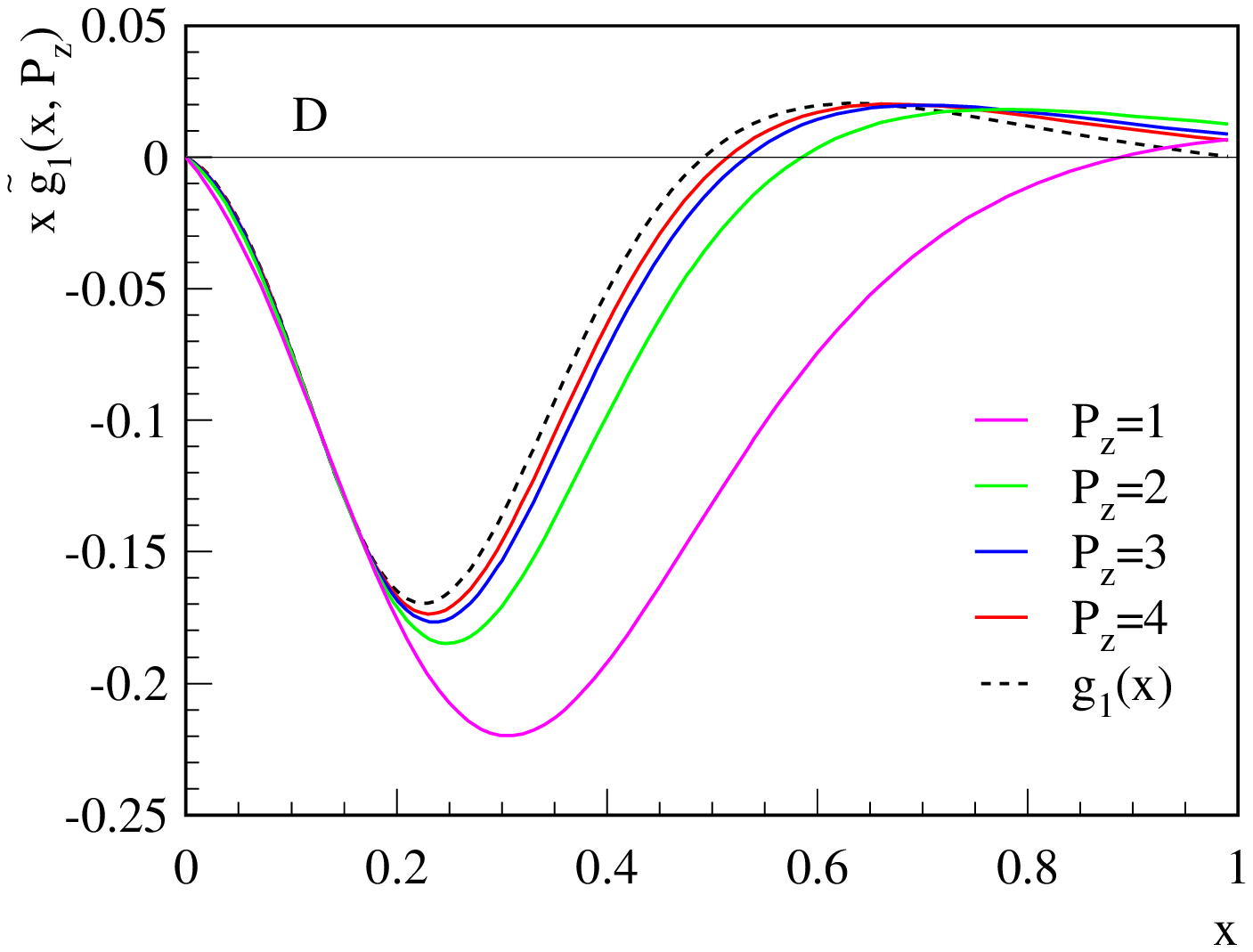, width=2.6in}
\caption{The helicity quasi-PDFs $x \tilde g_1(x, P_z)$ are plotted as a function of $x$ for $u$ (left) and $d$ (right) quark, respectively. Different lines are shown for $P_z=1$ GeV (purple), 2 GeV (green), 3 GeV (blue), and 4 GeV (red), respectively. The standard helicity distribution $g_1(x)$ (black dashed) is also shown for comparison.}
\label{fig:g1-x}
\eef

In Fig.~\ref{fig:f1-x}, we plot the quasi-unpolarized distribution $x\tilde f_1(x, P_z)$ as a function of momentum fraction $x$ for both up quark (left panel) and down quark (right panel) at different values of $P_z $;  1~GeV (purple), 2~GeV (green), 3~GeV (blue), and 4~GeV (red), respectively. For comparison, the standard unpolarized distribution $x f_1(x)$ is also shown (black dashed curve). It is important to realize that the quasi-PDFs have support for $-\infty<x<+\infty$~\cite{Ji:2013dva,Ma:2014jla,Xiong:2013bka}, and thus quasi-PDFs do not vanish for $x > 1$ at finite $P_z$. This is clearly seen in the figures: while $f_1(x)\to 0$ as $x\to 1$ for both $u$ and $d$ quarks,  at finite $P_z$, $\tilde f_1(x, P_z)$ remains finite when $x\to 1$.  It is evident that $\tilde f_1(x, P_z)$ has different behavior as compared with the standard distribution $f_1(x)$ for relatively small $P_z = 1$ GeV, as shown by the purple curves in Fig.~\ref{fig:f1-x}. However, once one increases $P_z \geq 2$ GeV, the shape of the quasi-PDFs approaches those of  the standard PDFs. 

In Figs.~\ref{fig:g1-x} 
we plot the quasi-helicity distribution $x\tilde g_1(x, P_z)$. We find very similar features to the unpolarized case. 
 For small $P_z=1$ GeV, the quasi-PDFs are different from the standard PDFs, but again, increasing $P_z\geq 2$ GeV, they become similar to the standard PDFs. Transversity distributions can be found in our paper~\cite{Gamberg:2014zwa}.
To further study the relative difference between quasi-PDFs and standard PDFs  quantitatively, we define the following ratios:
\bea
R_{f}^q(x, P_z)  =  \frac{\tilde f_1^q(x, P_z)}{f_1^q(x)},
\qquad
R_{g}^q(x, P_z)  =  \frac{\tilde g_1^q(x, P_z)}{g_1^q(x)},
\qquad
R_{h}^q(x, P_z)  =  \frac{\tilde h_1^q(x, P_z)}{h_1^q(x)},
\eea
As can be seen above for the intermediate $0.1\lesssim x\lesssim 0.4-0.5$ all the quasi-PDFs approximate  the corresponding standard PDFs to 
within $20-30\%$ when $P_z \gtrsim 1.5 - 2$ GeV, which seems within reach of lattice QCD calculations~\cite{Lin:2014zya}. 
On the other hand, as we have emphasized in last section, for the very large $x\sim 1$ region, the quasi-PDFs could be quite different from standard PDFs. This has already been demonstrated in Figs.~\ref{fig:f1-x}, \ref{fig:g1-x},  where the quasi-PDFs are still finite but the standard PDFs all vanish when $x\to 1$. Let us further make this point. In Fig.~\ref{fig:ratio-largex}, we plot the ratio $R^q(x, P_z)$ at large $x=0.7$ as a function of $P_z$ for $f_1^u$ (red), $f_1^d$ (blue), $g_1^u$ (green), and $h_1^d$ (purple), respectively. One can see that at $P_z \sim 1-2$ GeV, the ratio can be as large as $6-7$; that is,  in the large $x$ kinematics regime, the
 quasi-PDFs are quite different from the standard PDFs. In this kinematic regime,
one has to go to very large $P_z > 4$ GeV at least to obtain a good approximation to the standard PDFs. 
\bef
\psfig{file=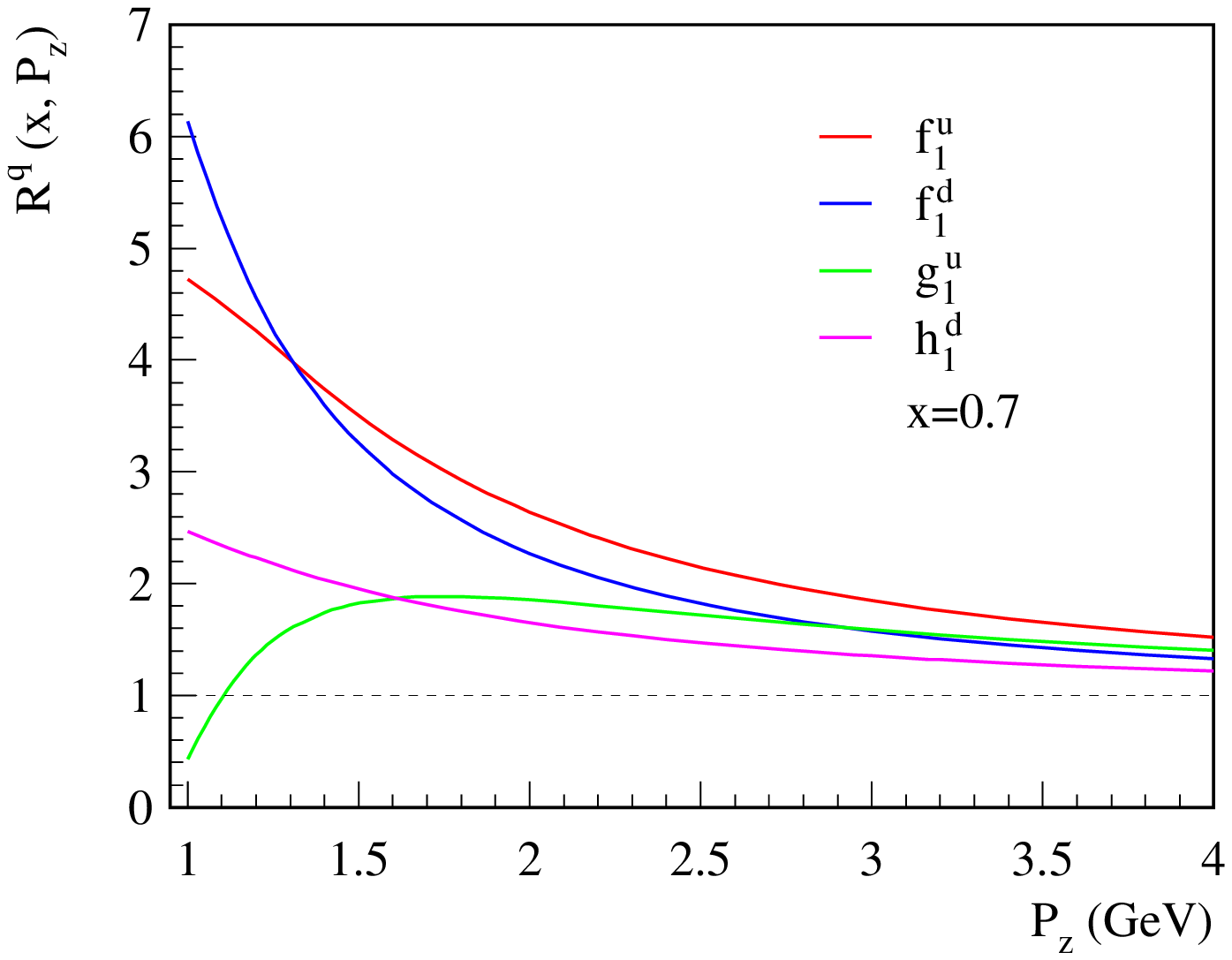, width=2.8in}
\caption{The ratio $R^q(x, P_z)$ at large $x=0.7$ as a function of $P_z$ for $f_1^u$ (red), $f_1^d$ (blue), $g_1^u$ (green), and $h_1^d$ (purple), respectively.}
\label{fig:ratio-largex}
\eef

\subsection{Positivity bound: Soffer inequality}

The Soffer inequality~\cite{Soffer:1994ww} relates the three leading-twist collinear PDFs $f_1,~g_1$,  and $h_1$ as 
\be
|h_1^{q}(x)| {\leq} \frac{1}{2}\left(f_1^{q}(x) + g_1^{q}(x) \right).
\ee
To test such an inequality for both standard PDFs and quasi-PDFs, let us define the following quantities:
\begin{eqnarray}
\tilde S^q(x, P_z) & = & \frac{1}{2}\left(f_1^{q}(x, P_z) + g_1^{q}(x, P_z)\right) - \left|h_1^{q}(x, P_z)\right|,
\\
S^q(x) & = & \frac{1}{2}\left(f_1^{q}(x) + g_1^{q}(x)\right) - \left|h_1^{q}(x)\right|.
\end{eqnarray}
The Soffer bound holds for the standard PDFs, thus we have $S^q(x) \geq 0$.

\bef
\psfig{file=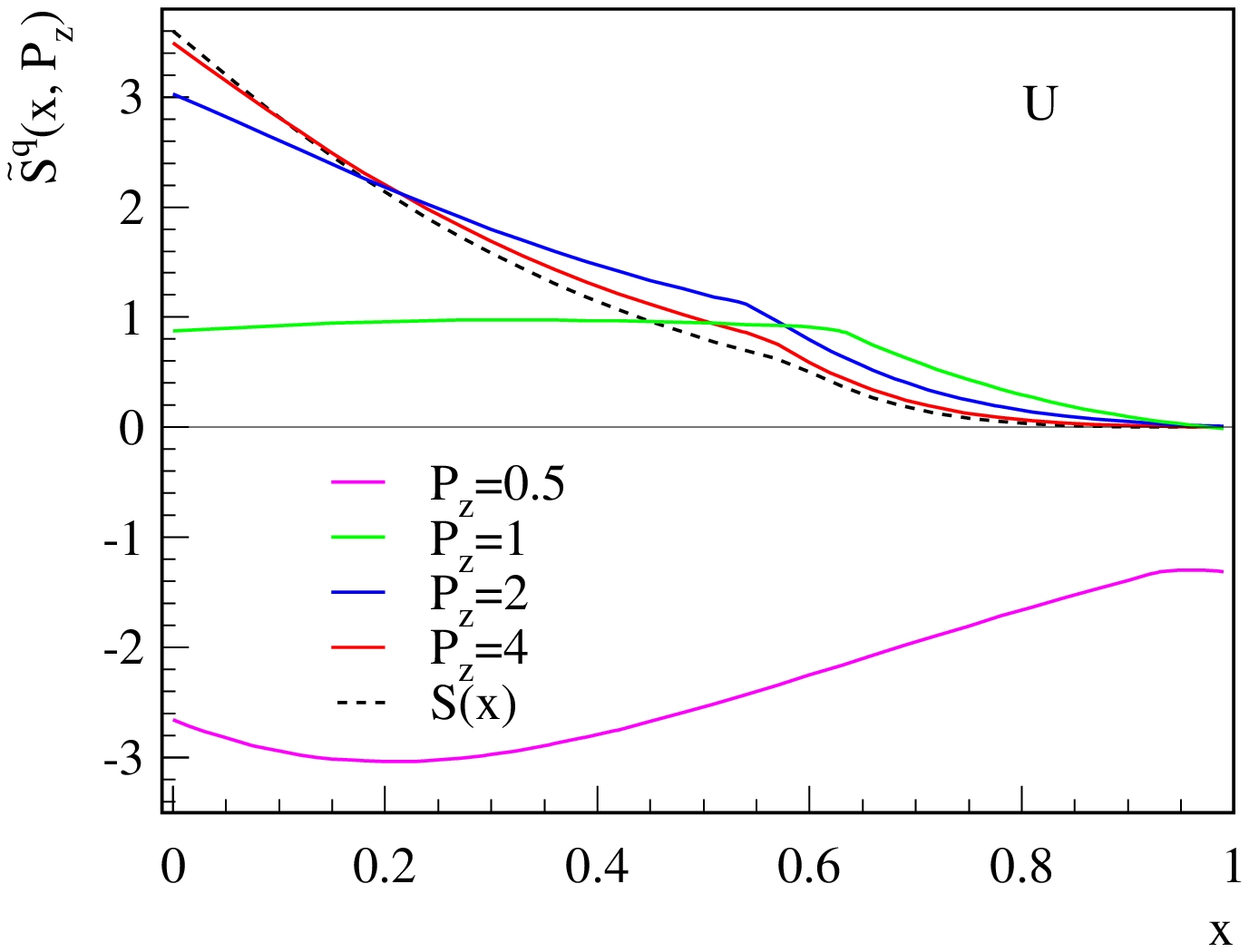, width=2.6in}
\hskip 0.2in
\psfig{file=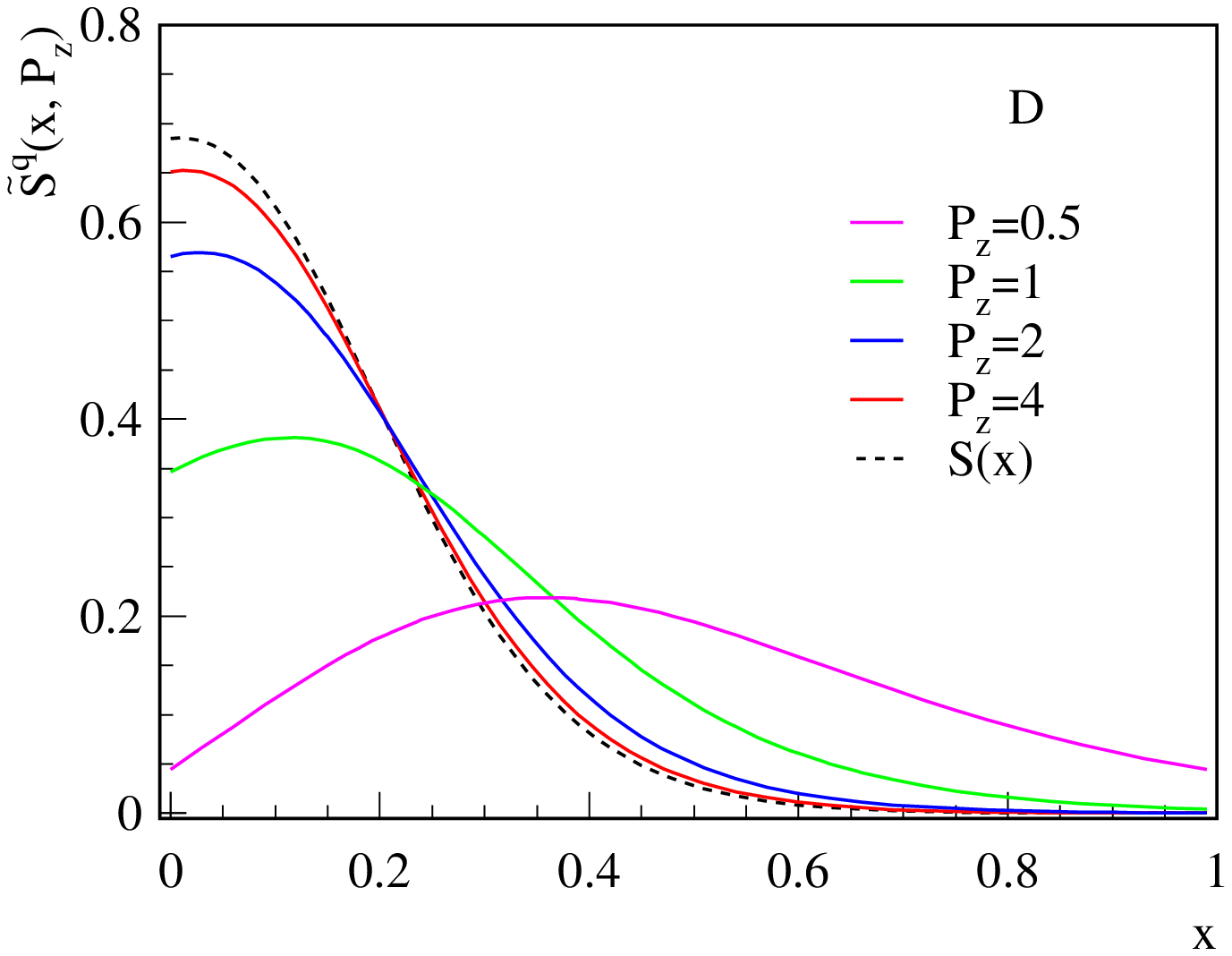, width=2.6in}
\caption{The function $\tilde S^q(x, P_z)$ is plotted versus $x$ for $u$ (left) and $d$ (right) quark. Different lines are shown for $P_z=0.5$ GeV (purple), 1 GeV (green), 2 GeV (blue),  4 GeV (red), respectively. The function $S^q(x)$ (black dashed) is also shown for comparison.}
\label{fig:soffer}
\eef
In Fig.~\ref{fig:soffer}, $\tilde S^q(x, P_z)$ is plotted versus $x$ for $u$ (left) and $d$ (right) quark at different values of  $P_z$, $0.5$ GeV (purple), 1 GeV (green), 2 GeV (blue), and 4 GeV (red), respectively. The function $S^q(x)$ (black dashed) for the standard PDFs is also shown for comparison. As one can see clearly from the black dashed curves, the Soffer bound is indeed satisfied for the standard PDFs for both $u$ and $d$ quarks. At the same time, within our spectator diquark model, as shown in the right panel of Fig.~\ref{fig:soffer}, for all the selected $P_z$ values, $\tilde S^q(x, P_z) \geq 0$ for the $d$ quark, 
that is,  the Soffer bound appears  to be satisfied for the $d$-quark quasi-PDFs. On the other hand, 
as shown in the left panel of Fig.~\ref{fig:soffer} for the $u$ quark, even though $\tilde S^q(x, P_z) \geq 0$ for $P_z=1,~2$, and 4 GeV,  for $P_z = 0.5$ GeV, $\tilde S^q(x, P_z) < 0$ for the entire plotted $0\leq x\leq 1$ region. In other words, the Soffer bound breaks down for relatively small $P_z$ values for the $u$ quark. What this tells us for  the usual lattice QCD simulations is that while the standard PDFs might still satisfy the positivity bounds, such as Soffer bound on the lattice~\cite{Diehl:2005ev}, these positivity bounds in general do not hold for quasi-PDFs, and, thus, one should avoid using them in lattice simulations.

\section{Conclusions \label{sec5}}

We used the spectator diquark model to consistently compare the quasi-parton distribution functions (PDFs) and the standard PDFs.  We took into account both the scalar diquark and axial-vector diquark contributions and generated  all the three leading-twist collinear PDFs, the unpolarized distribution~$f_1$, 
the helicity distribution~$g_1$, and the transversity distribution~$h_1$. Using the model parameters which lead to a reasonable description of the standard PDFs~$f_1^{u,d}(x)$ and~$g_1^{u,d}(x)$, consistent with those extracted from the global analysis~\cite{Bacchetta:2008af}, we presented  numerical studies for all quasi-PDFs. We found that for intermediate $0.1\lesssim x \lesssim 0.4-0.5$, the quasi-PDFs are good approximations for the corresponding standard PDFs  when the proton momentum $P_z \gtrsim 1.5-2$ GeV. However, in the large $x\sim 1$ region, a much larger $P_z > 4$~GeV  is necessary to obtain a similar accuracy of the approximation. By studying the Soffer positivity bound we found that the positivity bounds do not hold in general for the quasi-PDFs. Our study provides useful guidance for the lattice QCD calculations regarding the proton boost and accuracy of the quasi-PDFs approximation.

\section*{Acknowledgements}
This work is supported by the U.S. Department of Energy under Contract Nos. DE-FG02-07ER41460 (L.G.) and DE-AC02-05CH11231 (Z.K., I.V. and H.X.), and in part by the LDRD program at LANL.




\end{document}